\documentclass[prl,showpacs,preprintnumbers,twocolumn]{revtex4}
\usepackage{graphicx}
\usepackage{dcolumn}
\newcommand{\cupzclo}{Cu(pz)$_2$(ClO$_4$)$_2$}
\begin{document}
\bibliographystyle{apsrev}
\title{Quantum effects in a weakly-frustrated S=1/2 two-dimensional
Heisenberg antiferromagnet in an applied magnetic field}
\author{N. Tsyrulin$^{1,2}$, T. Pardini$^{3}$, R.~R.~P. Singh$^{3}$,
F. Xiao$^{4}$, P. Link$^{5}$, A. Schneidewind$^{5,6}$, A. Hiess$^{7}$,
C.~P. Landee$^{4}$, M.~M. Turnbull$^{8}$, M. Kenzelmann$^{1,9}$}

\affiliation{(1) Laboratory for Solid State Physics, ETH
Zurich, CH-8093 Zurich, Switzerland\\(2) Laboratory for
Neutron Scattering, ETH Zurich $\&$ Paul Scherrer Institute,
CH-5232 Villigen, Switzerland\\(3) Department of Physics, University of California,
Davis, Davis, California 95616, USA\\(4) Department of Physics, Clark University, Worcester, Massachusetts 01610,
USA\\(5)Forschungsneutronenquelle Heinz Maier-Leibnitz (FRM II),
D-85747 Garching, Germany\\(6)Institut f\"{u}r Festk\"{o}rperphysik,
TU Dresden, D-01062 Dresden, Germany \\(7)Institut Laue-Langevin,
Bo\^{\i}te Postale 156, F-38042 Grenoble, France\\(8)Carlson School
of Chemistry and Biochemistry, Clark University, Worcester,
Massachusetts 01610, USA\\(9) Laboratory for Developments and Methods,
Paul Scherrer Institute, CH-5232 Villigen, Switzerland}

\begin{abstract}
We have studied the two-dimensional S=1/2 square-lattice
antiferromagnet \cupzclo \ using neutron inelastic scattering and
series expansion calculations. We show that the presence of
antiferromagnetic next-nearest neighbor interactions enhances
quantum fluctuations associated with resonating valence bonds.
Intermediate magnetic fields lead to a selective tuning of
resonating valence bonds and a spectacular inversion of the
zone-boundary dispersion, providing novel insight into 2D
antiferromagnetism in the quantum limit.
\end{abstract}

\maketitle Some of the best examples of macroscopic quantum magnets
can be found in low-dimensional antiferromagnets that preserve
strong quantum fluctuations to very low temperatures. Cases in point
are antiferromagnetic (AF) chains that are quantum critical at zero
temperature and feature deconfined spinon excitations for S=1/2
\cite{Faddeev}, but a spin liquid with gapped triplet excitations
for S=1 \cite{Haldane}. In two dimensions, quantum effects are
generally reduced and the ground state of S=1/2 square lattice
Heisenberg antiferromagnets is long-range ordered, albeit with a
reduced ordered moment and spin-waves whose energy is renormalized
by quantum fluctuations \cite{Sandvik-qmc,Zheng}. Importantly,
however, short range quantum correlations are preserved in the
ground state and lead to a quantum-induced dispersion at the AF zone
boundary and to a weak continuum of states at high energies
\cite{Singh,Sandvik,Zheng,Ho,Christensen,McMorrow,Ronnow,Kim,Lumsden}.
\par
Competing interactions generally increase quantum fluctuations,
particularly for materials close to a quantum critical point. For
the 2D S=1/2 antiferromagnet, one source of competition can be
next-nearest neighbor interactions that destabilize a simple
nearest-neighbor antiparallel alignment and thus enhance
fluctuations \cite{Chandra,Read}. Such a scenario has been
extensively investigated using the $J_1 - J_2$ model, where $J_1$
and $J_2$ give the strength of the nearest and next-nearest neighbor
interactions \cite{Chandra}. The $J_1 - J_2$ model features two
different types of order for $J_2/J_1<0.38$ and $J_2/J_1>0.6$, and
probably a spin-liquid phase in between. The behavior of the 2D
S=1/2 antiferromagnet in magnetic fields was studied theoretically
for the case of nearest neighbor interactions
\cite{Zhitomirsky1,Zhitomirsky2,Syljuasen} as well as for $J_1 -
J_2$ model \cite{Schmidt, Thalmeier}. However, little is known
experimentally about the magnetic field behavior of the 2D S=1/2
antiferromagnet, particularly in the presence of next-nearest
neighbor interactions.
\par
We have studied a weakly-frustrated S=1/2 AF square-lattice where
next-nearest neighbor interactions are a small perturbation that
allows us to study the change of the magnetic properties upon
introduction of competing interactions. Our inelastic neutron
scattering experiments were performed using the 2D square lattice
Heisenberg antiferromagnet \cupzclo. Deuterated \cupzclo \
crystallizes in the monoclinic C2/c space group \cite{Woodward}.
Each ${\rm Cu^{2+}}$ ion in \cupzclo \ carries S=1/2 and is
surrounded by two pairs of identical cis pyrazine molecules,
creating a two dimensional square array of copper atoms linked by
pyrazine molecules in the $bc$ plane. The two fold rotation axis (0,
y, 1/2) and the mirror plane parallel to the $ac$ plane ensures
that all nearest neighbor interactions between $\rm Cu^{2+}$ ions on
the square lattice are identical. From the temperature dependence of
the magnetic susceptibility, it was concluded that \cupzclo \
represents a S=1/2 2D square lattice antiferromagnet with
nearest-neighbor exchange $J=1.53(8)\;\mathrm{meV}$ and a saturation
field of $\mu_0 H_{\rm sat}\sim 45\;\mathrm{T}$ \cite{Lancaster,
Turnbull}. Due to interplane interactions, \cupzclo \ adopts
long-range AF order below $T_{\rm N}=4.2\;\mathrm{K}$ and the ratio
between the interlayer to intralayer exchange has been estimated as
$J'/J=6.8\cdot10^{-4}$ \cite{Lancaster, Woodward}.\par
\begin{figure}
\includegraphics[width=0.45\textwidth, angle=0]{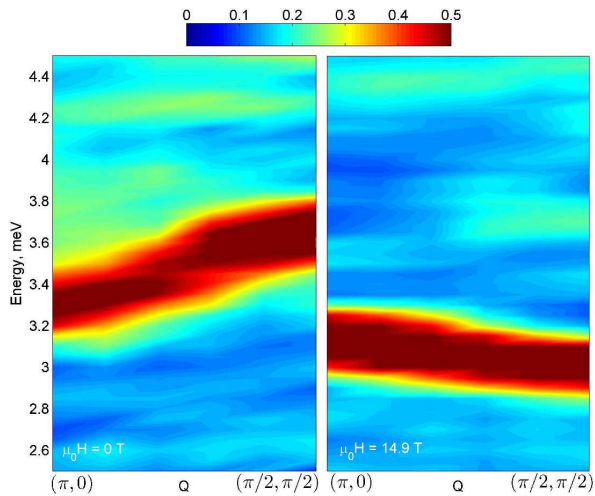}
  \caption{Zone boundary spin dispersion at zero field and ${\bf H \sim J}$.
  Color plot of the normalized scattering intensity $I(\textbf{Q},\omega)$ at
  $T=80\;\mathrm{mK}$, showing the dispersion from $(\pi,0)$ to $(\pi/2,\pi/2)$
  at zero field and $\mu_{0} H=14.9\;\mathrm{T}$ under otherwise identical conditions.
  Both panels present smoothed data obtained by performing six constant
  \textbf{Q}-scans from $(\pi/2,\pi/2)$ to $(\pi,\pi)$ with a $0.05\;\mathrm{meV}$ energy step.}
  \label{colorplot}
\end{figure}
The neutron experiments were performed using the cold-neutron
triple-axis spectrometers PANDA at FRM2 at Garching and IN14 at ILL
at Grenoble, and an array of deuterated \cupzclo \ single crystals
with a total mass up to 2g co-aligned with a mosaic of
$0.5^{\circ}$. The sample was aligned with its reciprocal (0, k, l)
plane in the horizontal scattering plane. A cryomagnet allowed the
application of vertical magnetic fields up to 14.9T. The magnetic
field was thus nearly perpendicular to the square-lattice plane. To
probe the ground state as a function of magnetic field, we used a
dilution refrigerator, reaching temperatures of the order of
50-100mK in each of the two experiments. The measurements were
performed using a fixed final energy $E_{\rm f}=4.66\;\mathrm{meV}$
and $E_{\rm f}=2.98\;\mathrm{meV}$ for the PANDA and IN14
experiments, respectively, obtained via the (002) Bragg reflection
from a pyrolithic graphite (PG) monochromator, a focused analyzer
and a cooled Be filter before the analyzer. The chemical unit cell
contains four ${\rm Cu^{2+}}$ atoms, thus nearest-neighbor AF order
does not break translational invariance, and $\textbf{Q}=(0,1,0)$
and $\textbf{Q}=(0,0,1)$ correspond to the AF point $(\pi,\pi)$.
Consequently, $\textbf{Q}=(0,m\pm 1/2,n\pm 1/2)$, where $m$ and $n$
are integers, correspond to $(\pi,0)$, while $\textbf{Q}=(0,m\pm
1/2,n)$ corresponds to $(\pi/2,\pi/2)$.
\par
A color plot of the normalized neutron scattering spectra at the
zone boundary is shown in Fig.~\ref{colorplot} for zero applied
field and $\mu_{0} H=14.9\;\mathrm{T}$. At zero field, the onset of
scattering at reciprocal wave-vector $(\pi,0)$ is reduced by
11.5(7)\% in energy compared to $(\pi/2,\pi/2)$, at odds with
spin-wave theory. This is larger than expected from Quantum Monte
Carlo simulations and series expansion calculations for the
Heisenberg square-lattice antiferromagnet with nearest-neighbor
interactions, which predict a zone-boundary dispersion of 8\% to
10\% \cite{Sandvik,Zheng}. Our observation is opposite to what has
been observed in the high-Tc material ${\rm La_2CuO_4}$ where the
energy at $(\pi,0)$ is higher than at $(\pi/2,\pi/2)$
\cite{Coldea-prl}. The latter has been attributed to ring-exchanges
arising from finite-$U/t$ \cite{Zheng-Hubbard}.
\par
\begin{figure}
\begin{center}
\includegraphics[height=4.3cm,bbllx=0,bblly=50,bburx=400,bbury=260,angle=0,clip=]{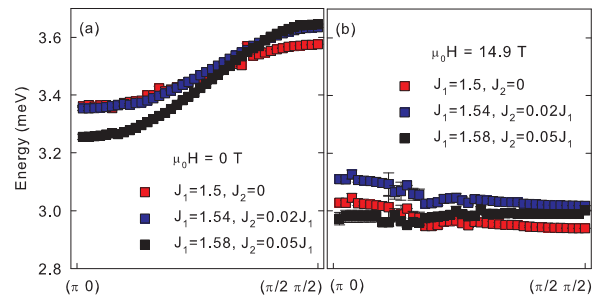}
  \caption{Series expansion calculations of the zone-boundary dispersion.
  Theoretical magnon dispersion for different values of the next-nearest
   neighbor exchange interaction $J_2$
  in zero magnetic field (left panel) and in a $\mu_{0} H=14.9\;\mathrm{T}$
  magnetic field (right panel). The value of the nearest neighbor interaction
   $J$ has been normalized to fit the experimental data. The normalization constants are shown in the legend.}
  \label{SE1}
\end{center}
\end{figure}
We developed series expansion calculations for the $J_1-J_2$ model
in a magnetic field \cite{advphys,book,zhe01}. Since the magnetic
field causes the spins to become non-collinear, this requires the
technically difficult multi-block method to be used \cite{zhe01}.
Our series expansion calculations for the $J_1-J_2$ model
\cite{advphys,book,zhe01} show that the addition of $J_2$ leads to
an enhancement of the zone boundary dispersion. Fig.~\ref{SE1} shows
the calculated single-magnon energies for different relative
strengths $J_2/J_1$ between nearest to next-nearest neighbor
interactions. For \cupzclo \ we thus attribute the increased
zone-boundary dispersion to a small AF next-nearest neighbor
interaction of the order of $J_2 \sim 0.02-0.05 J$.
\par
According to spin wave theory, the energy at $(\pi/2,\pi/2)$ is
equal to $E_{\rm (\pi/2,\pi/2)}=2 (J_1 - J_2)$ at all fields.
Next-nearest neighbor interactions thus lead to a smaller
zone-boundary energy, and thus to a smaller effective
nearest-neighbor exchange $J$. Using $J=J_1 - J_2$ and
$J=1.53(8)\;\mathrm{meV}$ determined from susceptibility
measurements \cite{Woodward,Lancaster}, the field-induced change of
$E_{\rm (\pi/2,\pi/2)}$ is obtained from the field dependence of the
renormalization factor $Z_{\rm c}=E_{\rm (\pi/2,\pi/2)}/2J$,
yielding $Z_{\rm c}=1.19(2)$ at zero field in excellent agreement
the predicted value of $Z_{\rm c}=1.18$ \cite{Sandvik}.
\par
\begin{figure}
\begin{center}
  \includegraphics[height=5.5cm,bbllx=0,bblly=0,bburx=495,bbury=315,angle=0,clip=]{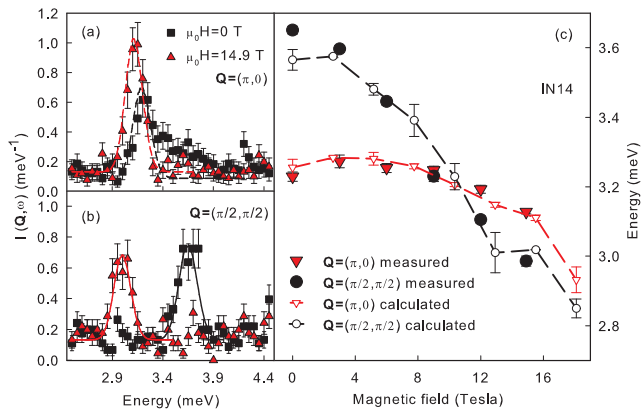}
  \caption{Field dependence of zone-boundary excitations.
  Energy scans at $(\pi,0)$ and $(\pi/2,\pi/2)$ are shown in
  (a) and (b) respectively. The data was measured at $T=80\;\mathrm{mK}$.
  The black solid line in (b) represents linear spin-wave excitations
  convoluted with the resolution function. Apart from the continuum region,
  all the experimentally measured peaks are resolution limited. The field dependent
  onset of magnetic scattering at $(\pi,0)$ and $(\pi/2,\pi/2)$ as a function of energy
  is shown in (c) by red triangles and black circles respectively. The dashed red and black
  lines represent the estimate from the series expansion calculations with
  $J_1=1.54\;\mathrm{meV}$ and $J_2=0.02J_1$.}
  \label{zbfield}
\end{center}
\end{figure}

Fig.~\ref{zbfield}a shows that the spin excitation at $(\pi,0)$ is
considerably broader than experimental resolution, and that neutron
scattering extends to about $E=3.9\;\mathrm{meV}$. This represents
clear experimental evidence of a magnetic continuum at $(\pi,0)$ of
the square-lattice antiferromagnet that is not expected from
spin-wave theory. The continuum is clearly stronger than in copper
deuteroformate tetradeuterate (CFDT) with a zone-boundary dispersion
of 7\% where next-nearest neighbor exchange are absent
\cite{McMorrow}, suggesting that  next-nearest neighbor exchange
enhances continuum excitations in the 2D S=1/2 square-lattice
antiferromagnet. The extended continuum in \cupzclo \ is also in
contrast to well-defined excitations observed in ${\rm La_2CuO_4}$
near $(\pi,0)$, suggesting that ring exchange has the opposite
effect on the continuum to next-nearest neighbor interactions.
Further, Fig.~\ref{colorplot} also provides evidence of an energy
gap around $(\pi/2,\pi/2)$ that separates a main mode and a much
weaker continuum above $4.2\;\mathrm{meV}$, as predicted by Ho {\it
et al.} \cite{Ho}.\par
Fig.~\ref{zbfield}c shows that magnetic fields strongly affect the
quantum fluctuations at the zone boundary: The energy of the
excitation at $(\pi/2,\pi/2)$ decreases much faster with
field than that at $(\pi,0)$. The zone-boundary dispersion at
$\mu_{0} H=14.9\;\mathrm{T}$ is inverted from that at zero
field. Our series expansion calculations show that, with the
application of a magnetic field, the inversion of the zone boundary
dispersion only occurs for sufficiently small $J_2/J_1$. A five
percent $J_2/J_1$ no longer shows the reversal of the dispersion
that is seen in experiments. This implies that even a relatively
small next-nearest neighbor interactions are effective in enhancing
the continuum of excitations, consistent with our estimate that
$J_2/J_1=0.02$. The quantum renormalization factor $Z_{\rm c}$
decreases rapidly from $Z_{\rm c}=1.19(2)$ at zero field and
approaches $Z_{\rm c}=0.99(2)$ at $\mu_{0} H=14.9\;\mathrm{T}$, also
consistent with our calculations.
\par
It is known that the wave-vector dependence of the magnetic
excitations at the zone boundary is a result of resonating valence
bond quantum fluctuations between nearest-neighbor spins that reduce
the energy of resonant excitations near $(\pi,0)$ below what is
expected from renormalized spin-wave theory \cite{Sandvik,McMorrow}.
The field-induced reversal of the zone-boundary dispersion reveals
that magnetic fields of the order of $H$ $\sim$ $J$ strongly couple to
these local quantum fluctuations. This dispersion, that is not
expected from spin-wave theory, demonstrates the presence of local
quantum fluctuations in the 2D S=1/2 square-lattice antiferromagnet
even for fields of the order of $H$ $\sim$ $J$. The quantum origin of
the dispersion reversal is also confirmed by our series expansion
calculations. The observed dispersion suggests that the energy of the
resonance at $(\pi,0)$ is raised by 4.5(7)\% above that of
renormalized spin-wave theory, providing direct evidence of
field-tuned resonating valence bond fluctuations.\par
\begin{figure}
\begin{center}
  \includegraphics[height=6.5cm,bbllx=0,bblly=0,bburx=376,
  bbury=311,angle=0,clip=]{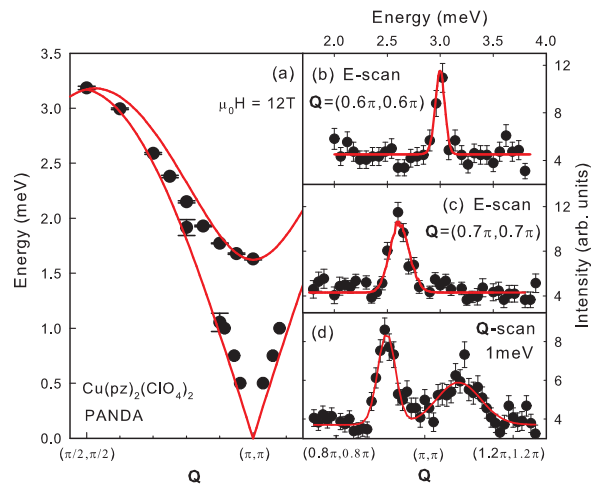}
  \caption{Spin dispersion for $H \sim J$. (a) The spin wave dispersion in
  \cupzclo \ measured at $\mu_{0} H=12\;\mathrm{T}$ and $T=80\;\mathrm{mK}$.
  The red line represents the spin-wave dispersion with $J_1=1.54\;\mathrm{meV}$
  and $J_2=0.02 J_1$. (b,c) Energy scans at $(0.6\pi,0.6\pi)$, $(0.7\pi,0.7\pi)$ provide evidence of a mode
with a field-induced gap. (d) Constant-energy scattering at
$1\;\mathrm{meV}$ energy transfer providing evidence
  for the Goldstone mode. The solid line in (b)-(d) corresponds to a convolution of a
  Gaussian with the resolution function, demonstrating that the excitations are resolution limited.}
  \label{12TDisp}
\end{center}
\end{figure}
\begin{figure}
\begin{center}
  \includegraphics[height=4.7cm,bbllx=0,bblly=207,bburx=290,
  bbury=380,angle=0,clip=]{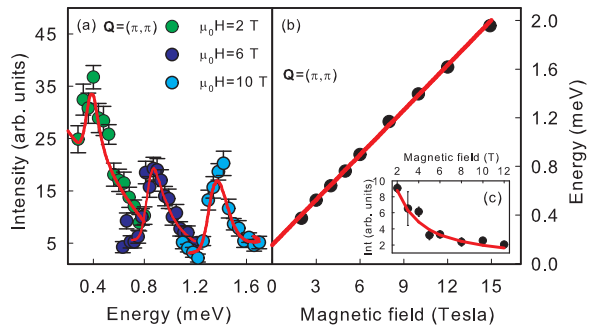}
  \caption{Field dependence of the zone-center excitation. Energy scans performed at
  $(\pi,\pi)$ at $\mu_{0} H=2\;\mathrm{T}$, $\mu_{0} H=6\;\mathrm{T}$,
  $\mu_{0} H=10\;\mathrm{T}$ (a) and the gap energy at the AF point plotted
  as the function of field (b). The data was measured  at $T=80\;\mathrm{mK}$.
  The solid lines in (a) are the fits of a Gaussian function convoluted with
  the resolution function. The filled circles in (b) represent the experimental
  data. The solid line in (b) is the linear fit of the gap energy. Black circles in inset (c)
  show the measured intensities as function of magnetic field and the curve
  is the scattering intensity calculated using linear spin-wave theory.}
  \label{GapvsField}
\end{center}
\end{figure}

The excitation spectrum at $\mu_{0} H=12\;\mathrm{T}$ features two
well defined magnetic modes of excitations (Fig.~\ref{12TDisp}).
The spectrum consists of a Goldstone mode, that indicates unbroken
rotational spin symmetry in the plane perpendicular to the magnetic
field, and a gapped mode. The field dependence of the mass of gapped
mode at the AF zone center, shown in Fig.~\ref{GapvsField}, is
linear, which is consistent with the theoretical arguments discussed
in \cite{Golosov}. However the finite gap at zero field, obtained by
the linear fit to the experimental data, implies the existence of a
small anisotropy in the system and requires additional
investigations. We also analyzed the spectral weight of the gapped
excitation at $(\pi,\pi)$. Our calculations completely describe the
observed field dependence of the peak intensities of the gapped mode
at the AF zone center (Fig.~\ref{GapvsField}c).\par
Fig.~\ref{12TDisp} shows the observed dispersion from $(\pi, \pi)$
to $(\pi/2, \pi/2)$ compared to linear spin-wave theory. The linear
spin-wave theory calculations in a magnetic field was performed for
the following Hamiltonian:
\begin{equation}
     \mathcal{H}=J_1\sum_{<i,j>}\textbf{S}_i\cdot\textbf{S}_j
     +J_2\sum_{<i,k>}\textbf{S}_i\cdot\textbf{S}_k
     -\mu_{\rm 0}H\sum_i\textbf{S}_i^{a},
\end{equation}where $\langle i,j\rangle$ represents nearest neighbor pairs of spins, $\langle i,k\rangle$ next nearest neighbor pairs and
$\textbf{S}^{a}$ is a spin component perpendicular to the $bc$
plane. The renormalized spin wave theory with $\rm{J_{eff}=1.54(1)meV}$ and $Z_{\rm c}=1.03(1)$ for $\mu_{0}H=12\;\mathrm{T}$ describes qualitatively the observed dispersion, as shown in Fig.~\ref{12TDisp}(a). However, for wave-vectors near $(3\pi/4,3\pi/4)$, spin-wave theory clearly predicts a larger spin-wave energy than experimentally observed. This difference may reflect the importance of magnon-magnon interactions \cite{Zhitomirsky1,Zhitomirsky2}. Possibly, better agreement could be obtained by including higher-order terms in the spin-wave calculation.\par

In summary, our study shows that the zone-boundary dispersion and
the zone-boundary continuum of the 2D S=1/2 square-lattice
antiferromagnet are enhanced by next-nearest neighbor interactions,
while the spin-wave energies merely experience a small
renormalization. Magnetic fields of the order of $H \sim J$ lead to
a qualitative change of the quantum fluctuations that suppress the
continuum of excitations and renormalize the spin-wave velocity, but
without suppressing the zone-boundary dispersion that arises from
non-trivial quantum fluctuations. In fact, in contrast to spin-wave
theory, we find that the zone-boundary dispersion is inverted
compared to zero field, providing direct evidence of a field-induced
change of resonating valence bond fluctuations.\par

\begin{acknowledgments}
We thank Dirk Etzdorf and Harald Schneider for technical assistance at PANDA spectrometer at FRM2
and the technical services of ILL involved during the experiment at IN14
This work was supported by the Swiss NSF (Contract No. PP002-102831).
\par
After submission of our manuscript, we became aware of A. Luscher
and A. Laeuchli \cite{Luscher} who also find theoretically that the
zone boundary dispersion inverts as a function of field.

\end{acknowledgments}

\newpage

\end{document}